\documentclass[aps,twocolumn,preprintnumbers,nofootinbib,floatfix]{revtex4-1}
\pdfoutput=1
\usepackage{dcolumn}
\usepackage{bm}
\usepackage{epsfig}
\usepackage{amsmath}
\usepackage{amssymb}
\usepackage{mathrsfs}
\usepackage{hyperref}
\usepackage[usenames,dvipsnames]{xcolor}
\usepackage{tikz-feynman}
\usepackage{graphicx} 
\usepackage{subfigure}
\tikzset{
    charged/.style={postaction={decorate},
        decoration={markings,mark=at position .55 with {\arrow[draw]{stealth}}}}
}

\def\be{\begin{equation}}
\def\ee{\end{equation}}
\def\bea{\begin{eqnarray}}
\def\eea{\end{eqnarray}}

\newcommand{\eqs}[1]{\begin{align} #1 \end{align}}
\newcommand{\mc}[1]{\mathcal{#1}}

\newcommand{\ket}[1]{| #1 \rangle}

\newcommand{\eq}[1]{\begin{equation}\begin{split} #1 \end{split}\end{equation}}

\newcommand{\Zs}[1]{}
\newcommand{\hc}{{\text{h.c.}}~}

\begin{document}

\title{\vspace*{-0.5cm}Entanglement Suppression, Enhanced Symmetry and\\ a Standard-Model-like Higgs Boson}

\author{\vspace{-0.2cm}
{Marcela Carena$^{\,1,2,3}$, Ian Low$^{\,4,5}$, Carlos E.~M.~Wagner$^{\,2,3,4}$ and Ming-Lei Xiao$^{\,4,5}$}
}
\affiliation{\vspace*{0.1cm}
$^1$\mbox{\small Fermi National Accelerator Laboratory, P.O. Box 500, Batavia, IL 60510, USA}\\
$^2$\mbox{\small Enrico Fermi Institute, University of Chicago, Chicago, IL 60637, USA}\\
$^3$\mbox{\small Kavli Institute for Cosmological Physics, University of Chicago, Chicago, IL 60637, USA}\\
$^4$\mbox{\small High Energy Physics Division, Argonne National Laboratory, Argonne, IL 60439, USA}\\
$^5$\mbox{\small Department of Physics and Astronomy, Northwestern University, Evanston, IL 60208, USA} \\
\vspace*{-0.5cm}
}

\begin{abstract}
We study  information-theoretic properties of  scalar models containing two Higgs doublets $\Phi_a$, where $a=1,2$ is the flavor quantum number. Considering the 2-to-2 scattering  $\Phi_a \Phi_b \to \Phi_c \Phi_d$ as a two-qubit system in the flavor subspace and the S-matrix as a quantum logic gate, we analyze the entanglement power of the S-matrix at the tree-level, in the limit the gauge coupling is turned off. Demanding the suppression of flavor entanglement during the scattering, the perturbative S-matrix in the broken phase can only be in the equivalent class of the Identity gate and the scalar potential exhibits a maximally enhanced $SO(8)$ symmetry acting on the 8 real components of the two doublets.  The $SO(8)$ symmetry leads to the alignment limit naturally, giving rise to a Standard-Model-like Higgs boson as a consequence of entanglement suppression.
\end{abstract}

\maketitle

\section{Introduction} 

The concept of symmetry is among the most powerful organizing principles in nature. However, very little has been said about its origin and whether symmetry can be derived from more fundamental principles. On the other hand, J.~A. Wheeler famously coined the phrase {\em It-from-bit}, stipulating that all things physical are information-theoretic in its origin \cite{Wheeler1989-WHEIPQ}. Given that entanglement is one of the most prominent features of quantum mechanics, one wonders if symmetry could arise out of a quantum information-theoretic origin.

Indeed, recent  studies in low-energy QCD revealed intriguing connections between the presence of emergent global symmetries and the  suppression of spin-entanglement in non-relativistic scattering of spin-1/2 baryons \cite{Beane:2018oxh,Low:2021ufv,Liu:2022grf}. Of particular interest is the interaction of neutron ($n$) and proton ($p$) in the low-energy, which exhibits an approximate $SU(4)_{sm}$ spin-flavor symmetry first observed by E.~P. Wigner \cite{Wigner449} more than half a century ago, and studied in modern perspective in Ref.~\cite{Mehen:1999qs}. Moreover, s-wave scattering lengths of $np$ are unusually large  in both the spin-singlet ($^1S_0$) and spin-triplet ($^3S_1$) channels, which are indicative of non-relativistic conformal invariance, also known as the Schr\"odinger symmetry \cite{Mehen:1999nd,Nishida:2007pj}. These emergent symmetries in the infrared  are usually characterized as fine-tuned or accidental. 

Ref.~\cite{Beane:2018oxh} first made the fascinating observation that, within the pionless effective field theory \cite{Kaplan:1998tg,Kaplan:1998we}, regions of parameter space where {  $SU(4)_{sm}$} and Schr\"odinger symmetry { emerge} coincide with regions where the spin-entanglement is suppressed in $np$ scattering. {In addition, entanglement suppression in  flavor-diagonal scattering of octet baryons leads to an even larger $SU(16)$ spin-flavor symmetry \cite{Beane:2018oxh}.}  Ref.~\cite{Low:2021ufv} studied these findings in an information-theoretic context and identified the association of the Identity gate with { $SU(4)_{sm}$ and $SU(16)$, as well as} the SWAP gate with the Schr\"odinger symmetry. These turned out to be the only two minimal entanglers for two qubit-systems \cite{Low:2021ufv}. Subsequently, Ref.~\cite{Liu:2022grf} extended the analysis to {flavor-changing} scattering of octet baryons and  identified scattering channels whose entanglement suppression are indicative of emergent $SU(6)$, $SO(8)$, $SU(8)$ and $SU(16)$ symmetries.

Given the nascent nature of this subject, it is important to proceed in an exploratory spirit and search for more examples of physical systems exhibiting a correlation between emergent symmetry and entanglement suppression. In this work we study a system of very different nature from the non-relativistic $np$ interaction: a  model of electroweak symmetry breaking containing two Higgs doublets $\Phi_a=(\Phi^+_a, \Phi^0_a)^T$, $a=1,2$, commonly referred to as two-Higgs-doublet models (2HDMs), which are the prototypical example for physics-beyond-the-Standard Model. We will analyze 2HDMs from a new perspective, focusing on   entanglement property of the S-matrix for the scattering  $\Phi_a \Phi_b \to \Phi_c \Phi_d$, in the limit the gauge coupling is turned off. The Yukawa coupling, on the other hand, allows us to define flavor quantum number of the Higgs doublets and does not contribute to tree amplitudes. We find that, in the broken phase, requiring the {\em perturbative} S-matrix to be a minimal entangler in the flavor space leads to a maximal $SO(8)$ symmetry, acting on the 8 real components of the two doublets. 

The  $SO(8)$ symmetry has an important consequence phenomenologically. Since measurements  at the Large Hadron Collider indicate properties of the 125 GeV Higgs is standard-model (SM) like \cite{ATLAS:2022vkf,CMS:2022dwd}, any viable 2HDMs must be in the ``alignment limit'' \cite{Gunion:2002zf,Carena:2013ooa,Carena:2014nza,Carena:2015moc,Low:2020iua}, where one of the CP-even mass eigenstates is SM-like. It turns out that imposing the $SO(8)$ symmetry  leads to ``natural alignment'' \cite{BhupalDev:2014bir,BhupalDev:2017txh}. Therefore entanglement suppression in 2HDMs gives rise to a SM-like Higgs boson.

{\section{Two-Qubit System Essentials}} 
\label{sect:2}
Here we briefly summarize the key concepts of quantum information needed for the present work. More comprehensive details can be found in Refs.~\cite{Low:2021ufv,Liu:2022grf}. 

We start with two distinguishable qubits, Alice ($A$) and Bob ($B$), each with its own basis vectors $\{|1\rangle_I, |2\rangle_I\}$, $I=A,B$. 
It is conventional to define the computation basis  $\{|11\rangle,|12\rangle,|21\rangle,|22\rangle\}$, where $|ij\rangle = |i\rangle_A\otimes |j\rangle_B$. There are several quantitative measures of entanglement in two-qubit systems  \cite{Kraus_2001}, although Ref.~\cite{Low:2021ufv} showed that all of them  are related to the concurrence $\Delta$ \cite{PhysRevLett.78.5022,PhysRevLett.80.2245}, which for a normalized state $|\Psi\rangle= c_{11} |11\rangle + c_{21} |21\rangle + c_{12}|12\rangle + c_{22} |22\rangle$ is defined as  
\be
\label{eq:conc}
\Delta(\Psi)=2|c_{11} c_{22} - c_{12} c_{21}| \ .
\ee 
The concurrence has a minimum at 0, if $|\Psi\rangle$ is not entangled,  and a maximum at 1 if it is maximally entangled. Other commonly employed entanglement measures include the von Neumann entropy $E_{vN}({\rho}) = -\mathrm{Tr}({\rho}_A\ln{\rho}_A)$ and the linear entropy $E_L({\rho}) = -\mathrm{Tr}[{\rho}_A({\rho}_A-1)]$, where ${\rho}=|\Psi\rangle\langle\Psi|$ is the density matrix and ${\rho}_{A/B}={\rm Tr}_{B/A}( {\rho})$ is the reduced density matrix for Alice/Bob.

Entanglement is a property of quantum states. Nevertheless we are more interested in the ability of a quantum operator $U$ to generate entanglement. In this regard, the entanglement power of a unitary operator is defined  by averaging over all direct product states that $U$ acts upon \cite{PhysRevA.63.040304,BallardWu2011}:
	\be\label{epdef}
	\Delta(U) = \overline{\Delta(U \left| \psi_A \right>\otimes\left| \psi_B \right>)} \ ,
	\ee
where the average is over each Bloch sphere.  Importantly, local operators which can be written as the product of singlet-qubit quantum gates, $V=U_A\otimes U_B$  do not generate entanglement. This defines an equivalent class among the two-qubit  gates, 
\be
\label{eq:singlqu}
U\sim U^\prime \ , \qquad {\rm if} \quad U=V_1 U^\prime V_2\ .
\ee
Operators in the same equivalent class have the same entanglement power.
Classification of all non-local, and hence entanglement generating operators in a two-qubit system has been achieved long  ago \cite{2000quant.ph.10100K,Kraus_2001,Zhang_2003}. However, for our purpose we focus on entanglement suppressing operators, which consist of only the Identity gate and the SWAP gate \cite{Low:2021ufv}, as well as their equivalent classes. In the computational basis they are defined by $\bm{1}\,|ij\rangle = |ij\rangle$ and  $\mathrm{SWAP}\,|ij\rangle = |ji\rangle$.
We will represent equivalent classes of $\bm{1}$ and SWAP by $[\bm{1}]$ and [SWAP], respectively.

In low-energy QCD, non-relativistic $np$ scattering is dominated by the $s$-wave and  
 the  S-matrix can be written as \cite{Low:2021ufv}
	\be
	\label{eq:smatrixnp}
	{S} 
	=  \frac{1}2\left(e^{2i\delta_1}+e^{2i\delta_0}\right) {\bm{1}} +\frac12\left(e^{2i\delta_1}-e^{2i\delta_0}\right) \text{ SWAP} \ ,
	\ee
where $\delta_{0}$ and $\delta_{1}$ are the scattering phases in the $^1S_0$ and $^3S_1$ channels, respectively. Then, Alice and Bob are  the spin-1/2 proton and neutron, respectively. 
One can see from Eq.~(\ref{eq:smatrixnp}) that ${ S} \propto {\bm{1}}$ if $\delta_0=\delta_1$ and ${ S} \propto {\rm SWAP}$ if $|\delta_0-\delta_1|=\pi/2$. The observation in Ref.~\cite{Beane:2018oxh} is  that $\delta_0=\delta_1$ corresponds to Wigner's $SU(4)_{sm}$ spin-flavor symmetry \cite{Wigner449,Mehen:1999qs} and $|\delta_0-\delta_1|=\pi/2$ gives rise to the Schr\"{o}dinger symmetry \cite{Mehen:1999nd,Nishida:2007pj}. Both are emergent symmetries not present in the fundamental QCD Lagrangian.
~\\

\section{2HDM Essentials}
\label{sect:3}

In 2HDM there are two hypercharge-one, $SU(2)$ doublet fields $\Phi_a=(\Phi_a^+, \Phi_a^0)^T,  a =1, 2$, and the most general potential is given by, following the notation of  Ref.~\cite{Haber:1993an},
\begin{align}
\label{eq:2hdmpot}
{\cal V}&=m^2_1 \Phi_1^\dag \Phi_1 + m^2_2 \Phi_2^\dag \Phi_2  -\left[ m_{12}^2 \Phi_1^\dag \Phi_2 + {\rm h.c.}\right] \nonumber \\
&  +\frac{\lambda_1}{2} (\Phi_1^\dag \Phi_1)^2+ \frac{\lambda_2}{2}(\Phi_2^\dag \Phi_2)^2 +\lambda_3 (\Phi_1^\dag \Phi_1)(\Phi_2^\dag \Phi_2)   \nonumber \\
& +\lambda_4 (\Phi_1^\dag \Phi_2)(\Phi_2^\dag \Phi_1)+ \left [\frac{\lambda_5}{2} (\Phi_1^\dag \Phi_2)^2 +  \lambda_6 (\Phi_1^\dag \Phi_1)( \Phi_1^\dag \Phi_2)\right.\nonumber\\
&\left. - \lambda_7 (\Phi_2^\dag \Phi_2)( \Phi_1^\dag \Phi_2)  + {\rm h.c.} \right ]\ .
\end{align}
For simplicity we  assume  CP conservation and $\lambda_i$ are real parameters, although our results can be easily generalized to the CP-violating case. 
We also assume a  $U(1)_{em}$-preserving vacuum, leading to two scalar VEVs, $v_1$ and $v_2$, that are real and non-negative, with $(v_1^2+v_2^2)^{1/2}\equiv v= 246$ GeV.  We define $t_\beta = {v_2}/{v_1} \ge 0$, $0\le \beta \le \pi/2$, such that $c_\beta\equiv \cos\beta=v_1/v$ and $s_\beta\equiv\sin\beta=v_2/v$.

Before considering the couplings of the two Higgs doublets to fermions, the flavor quantum number of Higgs doublets is not well-defined.  This is because $\Phi_1$ and $\Phi_2$ have identical SM quantum numbers and one is free to redefine the scalar fields by a global $U(2)$ rotation of $\vec{\Phi}=(\Phi_1, \Phi_2)^T$, which leaves the scalar kinetic term invariant,  $\vec{\Phi} \to \vec{\Phi}^\prime= {\cal U}\, \vec{\Phi}$, ${\cal U}^\dagger {\cal U} = \mathbb{I}$. { Parameters appearing in Eq.~(\ref{eq:2hdmpot}) are not invariant under $U(2)$ rotations, whereas the potentials related by $U(2)$ rotations are physically equivalent.} One can remove the $U(2)$ redundancy by introducing couplings to fermions. That is, once Yukawa couplings are introduced,  flavor can be defined.
For example, in type II 2HDMs \cite{Carena:2002es,Branco:2011iw}, one  doublet couples to the up-type fermions while the other couples to the down-type fermions, thereby allowing us to distinguish the two doublets. 
Another choice of basis, which is convenient  for the alignment limit \cite{Carena:2013ooa,Carena:2014nza,Carena:2015moc,Low:2020iua}, is the Higgs basis \cite{Botella:1994cs}, defined by  $(H_1, H_2)$ with the property: $\langle H_1^0 \rangle = {v}/{\sqrt{2}}$ and $\langle H_2^0 \rangle = 0$.  In the Higgs basis the scalar potential is {the} same  as in Eq.~(\ref{eq:2hdmpot}) but with the coefficients $\{m_1^2, m_2^2, m_{12}^2\}\to \{Y_1, Y_2, Y_3\}$ and $\lambda_i\to Z_i$. The minimziation of scalar potential gives  $Y_1 = - Z_1 v^2/2$ and ${Y}_3 = - {Z}_6 v^2/2$. { Making a $U(2)$ rotation corresponds to single-qubit operation and will not change the entanglement power of the S-matrix.}

The alignment limit, defined as when the scalar $h\equiv {\rm Re}(H_1^0)$, coincides with the 125 GeV mass eigenstate. In this case $h$, which carries the full VEV, couples to the massive gauge bosons with the SM strength when the gauge coupling is turned on.  It  is  shown in Refs.~\cite{Carena:2013ooa, Carena:2014nza,Carena:2015moc,Low:2020iua} that the alignment is achieved by the condition,
\be
Z_6 = 0 \ . 
\label{zeesix}
\ee
In the case of CP-violation, $Z_6$ is complex and the alignment condition is really two equations: ${\rm Re}(Z_6)=0$ and ${\rm Im}(Z_6)=0$, which eliminate mass mixings of $h$ with the other two neutral scalars. In any case, when $|Z_6|\ll 1$, $h$ is approximately aligned with the 125 GeV mass eigenstate, which then becomes SM-like. Moreover, in this limit $Z_1v^2$ is the dominant contribution to the mass of $h$: { $M_h^2 \approx Z_1 v^2=-2Y_1$}. To summarize, the mass of the SM-like Higgs boson is controlled by $Z_1$ while the departure from Higgs alignment is given by $Z_6$.

\section{S-matrix as an Identity gate}
\label{sect:entsym}

We now investigate the information-theoretic properties of 2HDMs, focusing on the S-matrix as an entanglement operator in the flavor-space in the scattering $\Phi_a\Phi_b\to \Phi_c\Phi_d$. In terms of Alice and Bob qubits, we identify $|i\rangle_A=\Phi^+_i$and $|i\rangle_B=\Phi^0_i$, $i=1,2$, respectively. The reason for choosing different electroweak quantum numbers is that Alice and Bob {are then associated with} distinguishable qubits. The S-matrix, being a unitary operator, then can be thought of as a two-qubit quantum logic gate. Recall that the S-matrix  is related to the transition matrix $T$: 
\be
\label{eq:Smat}
S= 1 + i \,T\ ,
\ee
where the matrix elements of the  T-matrix are given by 
\bea
\label{eq:ampT}
&& \langle \Phi_c\Phi_d |\, iT\, |\Phi_a\Phi_b\rangle \nonumber\\
&& \qquad = i (2\pi)^4 \delta^{(4)}(p_a+p_b-p_c-p_c) M_{ab,cd} \ .
\eea
$M_{ab,cd}$ are the scattering amplitudes one typically computes in perturbation  theory. Notice that the T-matrix, and therefore the amplitude itself, is not a unitary operator and does not admit an interpretation as a quantum gate. In fact,  unitarity of the $S$-matrix requires
\be
\label{eq:Tmatrix}
i(T^\dagger - T) =TT^\dagger\ ,
\ee
which is nothing but the  optical theorem. At the tree-level, the amplitude does not have an imaginary part and the T-matrix is Hermitian. This can be seen from the fact that, if $T\sim {\cal O}(\lambda)$ in perturbation, $T^\dagger T \sim {\cal O}(\lambda^2)$ is higher order in the coupling constants and the right-hand side of Eq.~(\ref{eq:Tmatrix}) can be ignored;  perturbative unitarity of the S-matrix is fulfilled at ${\cal O}(\lambda)$. It is worth pointing out that our approach is different from some in the literature which looked at the entanglement property of the amplitude, instead of the S-matrix \cite{Cervera-Lierta:2017tdt,Miller:2023ujx,Bai:2023tey}.

We are interested in an S-matrix which suppresses flavor entanglement in 2-to-2 scattering, when turning off the gauge fields.     A priori we need to consider the two equivalent classes associated with the Identity  and the SWAP gates, $[\bm{1}]$ and [SWAP], {respectively}~\cite{Low:2021ufv}. However, we argue that perturbatively the S-matrix could only be in $[\bm{1}]$ and not [SWAP]. This is  most clear by looking at Eq.~(\ref{eq:Smat}), which implies
\begin{align}
S &\sim  [\bm{1}] & \Leftrightarrow & &T &\sim [\bm{1}]\ ,\\
S &\sim  {\rm [SWAP]} & \Leftrightarrow&  &T &\sim i\,{ ( [\bm{1}] + {\rm [SWAP]} )}  \ .
\end{align}
In other words, the S-matrix being in [SWAP] requires a {tree-level} cancellation between the T-matrix, which we compute in perturbation, against the non-interacting part of the S-matrix. This can only be achieved in a strongly-coupled theory. Indeed, Refs.~\cite{Mehen:1999nd,Nishida:2007pj} found the SWAP gate is associated with fermionic systems interacting with the largest strength allowed by unitarity -- fermions at unitarity -- and Schr\"{o}dinger symmetry emerges from it.  For weakly coupled theories, an entanglement-suppressing S-matrix can only be in $[\bm{1}]$ at finite orders in perturbation theory, except when a certain class of diagrams is resummed to all orders. An example is the eikonal limit where the $t$-channel diagram dominates in the 2-to-2 scattering and exponentiates, in which case the S-matrix is  manifestly unitary \cite{Aoude:2020mlg}.

In what follows we will focus on the flavor subspace of the amplitude $M_{ab,cd}$ defined in Eq.~(\ref{eq:ampT}),  which is Hermitian at the tree-level, and work out the conditions on the amplitude in order for the S-matrix to be in $[\bm{1}]$. Starting from an initial product state in the flavor space, {$|\Phi_a\Phi_b\rangle= (\kappa |1\rangle+ \epsilon |2\rangle)\otimes (\gamma |1\rangle+ \delta|2\rangle)$, where $|\kappa|^2+|\epsilon|^2=|\gamma|^2+|\delta|^2=1$. }The outgoing state then has the flavor structure,
\eq{
    \ket{\Phi_c\Phi_d} &= (\delta_{ac}\delta_{bd} + iM_{ab,cd}) \ket{\Phi_a}\otimes\ket{\Phi_b} \ .
}
{Demanding that the concurrence in Eq.~(\ref{eq:conc}) vanishes, $\Delta( \ket{\Phi_c\Phi_d}) =0$}, and working to the first order in perturbation by keeping only terms linear  in $M_{ab,cd}$, we obtain 
\eqs{\label{eq:id_equiv1}
    & M_{11,11}+M_{22,22}=M_{12,12}+M_{21,21}\ ,\\
    & M_{11,22}=M_{12,21}=M_{21,12}=M_{22,11}=0\ ,\label{eq:id_equiv2}\\
    & M_{11,12}=M_{21,22}\ ,\quad M_{11,21}=M_{12,22}\ .\label{eq:id_equiv3}
}
More details can be found in the Supplemental Materials. These are the conditions the tree-level amplitude must satisfy in order for the S-matrix to be in the equivalent class of the Identity gate, $S=[\bm{1}]$, which are more general than simply requiring $M_{ab,cd} = \bm{1}$. This situation is markedly different from that in the $np$ scattering, where rotational invariance  constrains  the $s$-wave S-matrix to be exactly $\bm{1}$ perturbatively. If we had imposed the $SU(2)$ flavor symmetry in our 2HDMs, we would have arrived at the same situation.

\section{Emergent $SO(8)$ Symmetry}

\begin{figure*}[t]
\centering
    \subfigure[]{
    \begin{tikzpicture}[scale=.9]
    \node (a) at (0,2) {$\Phi^+_a$};
    \node (b) at (0,0) {$\Phi^0_b$};
    \node (c) at (2,2) {$\Phi^+_c$};
    \node (d) at (2,0) {$\Phi^0_d$};
    \draw[charged] (a) -- (1,1);
    \draw[charged] (1,1) -- (c);
    \draw (b) -- (1,1) -- (d);
    \end{tikzpicture}
    }
    \hfill
    \subfigure[]{
    \begin{tikzpicture}[scale=.9]
    \node (a) at (0,2) {$\Phi^+_a$};
    \node (b) at (0,0) {$\Phi^0_b$};
    \node (c) at (2,2) {$\Phi^+_c$};
    \node (d) at (2,0) {$\Phi^0_d$};
    \coordinate (o1) at (0.4,1);
    \coordinate (o2) at (1.6,1);
    \draw[charged] (a) -- (o1);
    \draw[charged] (o1) --node[below]{$P_{s,i}$} (o2);
    \draw[charged] (o2) -- (c);
    \draw (b) -- (o1) -- (o2) -- (d);
    \end{tikzpicture}
    }
    \hfill
    \subfigure[]{
    \begin{tikzpicture}[scale=.9]
    \node (a) at (0,2) {$\Phi^+_a$};
    \node (b) at (0,0) {$\Phi^0_b$};
    \node (c) at (2,2) {$\Phi^+_c$};
    \node (d) at (2,0) {$\Phi^0_d$};
    \coordinate (o1) at (1,1.6);
    \coordinate (o2) at (1,0.4);
    \draw[charged] (a) -- (o1);
    \draw[charged] (o1) -- (c);
    \draw (o1) --node[left]{$P_{t,i}$} (o2);
    \draw (b) -- (o2) -- (d);
    \end{tikzpicture}
    }
    \hfill
    \subfigure[]{
    \begin{tikzpicture}[scale=.9]
    \node (a) at (0,2) {$\Phi^+_a$};
    \node (b) at (0,0) {$\Phi^0_b$};
    \node (c) at (2,2) {$\Phi^+_c$};
    \node (d) at (2,0) {$\Phi^0_d$};
    \coordinate (o1) at (1,1.6);
    \coordinate (o2) at (1,0.4);
    \draw[charged] (a) -- (o1);
    \draw[charged] (o2) -- (c);
    \draw[charged] (o1) --node[left]{$P_{u,i}$} (o2);
    \draw (b) -- (o2);
    \draw (o1)-- (d);
    \end{tikzpicture}
    }
    \caption{Feynman diagrams of $\Phi^+_a\Phi^0_b \to \Phi^+_c\Phi^0_d$ scattering in the symmetry broken phase.}\label{fig:1}
\end{figure*}
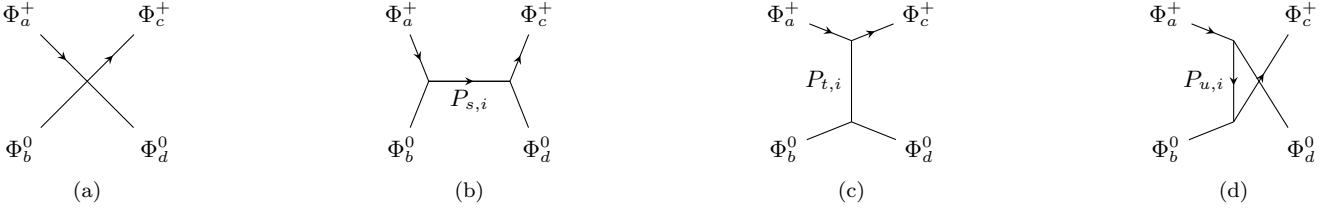

In this section we compute the tree-level scattering amplitude for $\Phi_a^+\Phi_b^0 \to \Phi_c^+\Phi_d^0$ in the broken phase. The goal is to demonstrate that, when the 2-to-2 amplitude minimizes entanglement and satisfies Eqs.~(\ref{eq:id_equiv1}-\ref{eq:id_equiv3}), a maximal $SO(8)$ symmetry emerges. The 2-to-2 amplitude includes four Feynman diagrams shown in Fig.~\ref{fig:1}: the 4-point contact interaction  and the $s/t/u$-channels mediated by  cubic vertices in the broken phase. The internal propagators in  Fig.~\ref{fig:1} necessitates a rotation into the mass eigenstates, which in general is different between the charged sector and the neutral sector. However, an advantage of the Higgs basis is that the charged sector is already diagonal since $t_\beta=0$. So we will perform that calculation in the Higgs basis, $H_1 =  (G^+, v/\sqrt{2}+H^0_1)^T$ and $H_2 = ( H^+, H^0_2)$, where $G^+$ is the charged Goldstone boson and $H^+$ is the massive charged scalar. In the neutral sector there are four mass eigenstates which we denote by $(h, H, G^0, A)$: $h$ is the lightest CP-even scalar, which we assume to be the 125 GeV Higgs boson, $H$ and $A={\rm Im}[H_2^0]$ are the CP-even and CP-odd heavy scalars, respectively, and $G^0={\rm Im}[H_1^0]$ is a Goldstone boson. The rotation matrix $\mc{R}$ in the neutral sector is given by
\eq{
 \!\!\!\!\!\!   \begin{pmatrix}
        h \\ H \\ G^0 \\ A
    \end{pmatrix} = \mc{R} \begin{pmatrix}
        H^0_1 \\ H^0_1{}^* \\ H^0_2 \\ H^0_2{}^*
    \end{pmatrix}, \
     \mc{R} = \frac{1}{2}\begin{pmatrix}
        -s_{\tilde{\alpha}} & -s_{\tilde{\alpha}} & c_{\tilde{\alpha}} & c_{\tilde{\alpha}} \\
        c_{\tilde{\alpha}} & c_{\tilde{\alpha}} & s_{\tilde{\alpha}} & s_{\tilde{\alpha}} \\
        -i & i & 0 & 0 \\
        0 & 0 & -i & i
    \end{pmatrix} ,
    }
where ${\tilde{\alpha}}$ is the mixing angle in the neutral CP-even sector in the Higgs basis. It is related to the corresponding mixing angle $\alpha$ in the general basis  by $\tilde{\alpha}=\alpha-\beta$. Observe that the alignment condition corresponds to $c_{\tilde{\alpha}}=0$.

The full amplitude is
\eqs{\label{eq:amp_broken}
iM_{ab,cd} &=  iM{}_{ab,cd}^0 - \frac{v^2}{2}\sum_{i}\sum_{r=s,t,u} M^r_i{}_{ab,cd}\ P_{r,i} \ , \\
\label{eq:m0abcd}
M^0_{ab,cd} &= \begin{pmatrix}
		Z_1	&	Z_6	&	Z_6	&	Z_5	\\
		Z_6	&	Z_3	&	Z_4	&	Z_7	\\
		Z_6	&	Z_4	&	Z_3	&	Z_7	\\
		Z_5	&	Z_7	&	Z_7	&	Z_2
	\end{pmatrix} \ , \\
 M^s_i{}_{ab,cd} &= M_{abi}M^*_{cdi}\ ,\quad M^u_{i\ ab,cd} = M_{adi}M^*_{cbi}\ ,\\
 M^t_i{}_{ab,cd} &= \sum_{j,k}\mc{R}_{ij}M_{ajc}(\mc{R}_{ik}M_{dkb,0})^* + \hc\ ,
 } 
where the propagators entering the $s/t/u$-channel diagrams are $P_{t,i} = i/(t-m_{0,i}^2)$ and $P_{r,i} = i/(r-m_{+,i}^2)$, for $r=s,u$.  Masses in the propagator run through $m_{0,i}=\{m_h, m_H, 0, m_A\}$ and $m_{+,i}=\{m_{H^\pm}, 0\}$. Moreover, the  cubic vertices $M_{dkb}$ and $M_{dkb,0}$ are
\eqs{
\!\!\! \!\!  &\!\!\!\left.\frac{\partial {\cal V}}{\partial v}\right|_{v=0}  = \frac{1}{\sqrt{2}} \sum_{a,b,c}\left[M_{abc}H_a^+H_b^0H_c^-\right. \nonumber \\
   &\qquad\qquad\qquad \left. + \frac{1}{2}M_{abc,0}H_a^0H_b^0H_c^0{}^* + \hc \right]\, .
}
In order for the S-matrix to minimize entanglement and be in $[\bm{1}]$ for arbitrary kinematics, we will demand that every term in Eq.~(\ref{eq:amp_broken}) satisfies the conditions in Eqs.~(\ref{eq:id_equiv1}-\ref{eq:id_equiv3}). For $M^0_{ab,cd}$ in Eq.~(\ref{eq:m0abcd}), they lead to $Z_1 + Z_2 = 2Z_3$, $Z_4 = Z_5 = 0$, and $Z_6 = Z_7$. These relations greatly simplify expressions in  $M^r_i{}_{ab,cd}$. Solving for entanglement suppressing amplitudes in the $s/t/u$-channel then requires: 
\eqs{
\label{eq:entsupp}
&Z_1=Z_2=Z_3\equiv {Z} \ ,& &Z_i= 0 \ ,\ \  i \neq 1,2,3  \ ,\\
\label{eq:entsupp2}
&Y_1=Y_2\equiv Y  = { -Z v^2/2} \ ,& &Y_3=0 \ ,
}
which lead to  the scalar potential, 
\eq{
\mc{V} &= Y (H_1^\dag H_1 + H_2^\dag H_2) + \frac{{Z}}{2}(H_1^\dag H_1 + H_2^\dag H_2)^2 \\
	 &= \frac{{Z}}{2}\left(\sum_{i=1,2} |H_i^0|^2 + G^+G^-+H^+H^- - \frac{v^2}2\right)^2, \label{eq:Massterm}
}
 (Complete Feynman rules and full expressions for the  amplitudes can be found in the Supplementary Material.) The above potential  exhibits a maximal $SO(8)$ symmetry acting on the 8 real components of the two  doublets and is spontaneously broken to $SO(7)$. 
The spectrum contains a massive scalar $h$ with $m_h^2 = -2 Y = Z v^2$, while  all other scalars are exact Goldstone bosons and massless.  However, recall that the $SO(8)$ symmetry is explicitly broken by  Yukawa and gauge couplings (when turned on) and the Goldstone bosons will either become massive or be ``eaten'' by the $W$ and $Z$ bosons.
Furthermore, to achieve a realistic mass spectrum consistent with  null searches at the LHC,   $SO(8)$  needs to be broken softly by the mass terms~\cite{BhupalDev:2017txh}. Since one of the minimization conditions  relates $Y_3$ to $Z_6$, which controls the alignment condition, one could leave $Y_3=0$ and introduce an additional $Y_2$ contribution,  which fixes the non-standard Higgs spectrum $m_H^2 = m_A^2 = m_{H^\pm}^2 = Y_2 + Z v^2/2$ (see, for example, Ref.~\cite{Bahl:2022lio}). The latter clearly shows that,  in the $SO(8)$ symmetric limit, Eq.~(\ref{eq:entsupp2}) leads to  massless  non-standard Higgs bosons.

\section{Conclusions}

In this work we analyzed information-theoretic properties of general 2HDMs, a prototypical example for beyond-the-SM theories.
When the gauge and Yukawa couplings  are turned off, demanding that the perturbative S-matrix suppresses flavor entanglement  in $\Phi_a\Phi_b\to \Phi_c\Phi_d$,  and is in the equivalent class of the Identity gate, singles out  regions of parameter space where the  $SU(2)\times U(1)$ symmetry is enhanced to $SO(8)$, which in turn is broken spontaneously to $SO(7)$ and gives rise to a SM-like Higgs  boson as a consequence of entanglement suppression.Turning on the Yukawa and gauge couplings results in explicit $SO(8)$ breaking and lifts the otherwise  massless non-SM Higgs bosons. However, a realistic  spectrum compatible with current LHC bounds requires the $SO(8)$ to be  further broken softly by  mass terms and the entanglement suppression is  approximate. We leave for future work a detailed analysis of the degree of entanglement suppression and its phenomenological implications for LHC data. To summarize, the unexpected connections between quantum entanglement, emerging symmetries and a SM-like Higgs boson provide  potentially rich and fruitful  insights for the  exploration of  physics beyond-the-SM using tools in quantum information science.

\section*{Acknowledgement} 

We thank R. Aoude and M. Savage for comments on the manuscript. Fermilab is operated by Fermi Research Alliance, LLC under Contract No. DE-AC02-07CH11359 with the U.S. Department of Energy. C.W.  is supported in part by the U.S. Department of Energy grant DE-SC0013642. Work at Argonne is supported in part by the U.S. Department of Energy under contract DE-AC02-06CH11357. I.L. is supported in part by the U.S. Department of Energy  grant  DE-SC0023522. M.-L.X. is supported in part by the Jay Jones Fund in the Department of Physics and Astronomy at Northwestern University.

\bibliography{Ent_2hdm_ref}

\onecolumngrid
\subsection*{\large\bf Supplementary Material for ``Entanglement Suppression, Enhanced Symmetry and a Standard-Model-like Higgs Boson''}

Derivation of Eqs.~(\ref{eq:id_equiv1}-\ref{eq:id_equiv3}): we assume the initial state is
\eq{
    \ket{\Phi_a} = \kappa\ket{1}+\epsilon\ket{2}\ ,\quad \ket{\Phi_b} = \gamma\ket{1}+\delta\ket{2}\ ,
}
where $|\kappa|^2+|\epsilon|^2=|\gamma|^2+|\delta|^2=1$.   The final state is 
\eq{
    \ket{\Phi_c\Phi_d} &= (\delta_{ac}\delta_{bd} + iM_{ab,cd}) \ket{\Phi_a}\otimes\ket{\Phi_b} = c_{ij}\ket{ij} \\
    c_{11} &= (1+iM_{11,11})\,\kappa\gamma + i M_{12,11}\ \kappa\delta+i M_{21,11}\ \epsilon\gamma + i M_{22,11}\ \epsilon\delta\ ,\\
    c_{12} &= i M_{11,12}\ \kappa\gamma + (1+i M_{12,12})\, \kappa\delta + i M_{21,12}\ \epsilon\gamma + i M_{22,12}\ \epsilon\delta\ ,\\
c_{21} &= i M_{11,21}\ \kappa\gamma + i M_{12,21}\ \kappa\delta +(1+ i M_{21,21})\,\epsilon\gamma + i M_{22,21}\ \epsilon\delta\ ,\\
c_{22} &= iM_{11,22}\ \kappa\gamma + i M_{12,22}\ \kappa\delta+i M_{21,22}\ \epsilon\gamma +(1+ i M_{22,22})\, \epsilon\delta\ ,
}
The concurrence $\Delta( \ket{\Phi_c\Phi_d}) = c_{11}c_{22} - c_{12}c_{21}$ reads
\eq{
     \Delta( \ket{\Phi_c\Phi_d}) &= i\kappa\epsilon\gamma\delta(M_{11,11}-M_{12,12}-M_{21,21}+M_{22,22}) \\
     &\quad +i \kappa\epsilon(\gamma^2-\delta^2)( M_{21,22}-M_{11,12})+i (\kappa^2-\epsilon^2)\gamma\delta (M_{12,22}-M_{11,21}) \\
    &\quad -i M_{12,21}\ \kappa^2\delta^2-i M_{21,12}\ \epsilon^2\gamma^2+i M_{11,22}\ \kappa^2\gamma^2+i M_{22,11}\ \epsilon^2\delta^2 + O((M_{ab,cd})^2)\ .
}
Since $\kappa, \epsilon, \gamma$ and $\delta$ are arbitrary, setting $\Delta( \ket{\Phi_c\Phi_d})=0$ leads to the conditions in Eqs.~(\ref{eq:id_equiv1}-\ref{eq:id_equiv3}).

In the Higgs basis, the minimization condition leads to the quadratic coefficients $Y_1=-Z_1v^2/2$ and $Y_3=-Z_6 v^2/2$, while the mass matrices of the charged and CP even/odd neutral scalars are given by
\eqs{\label{eq:masq}
	m^2_+ &= \begin{pmatrix} 0 & 0 \\ 0 & Y_2+Z_3v^2/2 \end{pmatrix} ,\\
	m^2_{\rm even} &= \begin{pmatrix} Z_1 v^2 & Z_6 v^2 \\ Z_6 v^2 & Y_2+(Z_3+Z_4+Z_5)v^2/2 \end{pmatrix} ,\\
	m^2_{\rm odd} &= \begin{pmatrix} 0 & 0 \\ 0 & Y_2+(Z_3+Z_4-Z_5)v^2/2 \end{pmatrix} \ .
}
The Feynman rules are given by (time goes from left to right)
\eq{
\begin{tikzpicture}[baseline=(a.base)]
\node (a) at (0,1) {$H^0_1$};
\node (b) at (1.5,2) {$H^+_a$};
\node (c) at (1.5,0) {$H^-_b$};
\draw (a) -- (1,1);
\draw[charged] (c) -- (1,1);
\draw[charged] (1,1) -- (b);
\end{tikzpicture} = \frac{iv}{\sqrt{2}}\begin{pmatrix}
	Z_{1}   &   Z_{6} \\
	Z_{6}   &   Z_{3}
\end{pmatrix}_{ab} \ ,}
\eq{
\begin{tikzpicture}[baseline=(a.base)]
\node (a) at (0,1) {$H^0_2$};
\node (b) at (1.5,2) {$H^+_a$};
\node (c) at (1.5,0) {$H^-_b$};
\draw (a) -- (1,1);
\draw[charged] (c) -- (1,1);
\draw[charged] (1,1) -- (b);
\end{tikzpicture} = \frac{iv}{\sqrt{2}}\begin{pmatrix}
	Z_{6}   &   Z_{5} \\
	Z_{4}   &   Z_{7}
\end{pmatrix}_{ab}\ , }
\eq{
\begin{tikzpicture}[baseline=(a.base)]
\node (a) at (0,1) {$H^0_1$};
\node (b) at (1.5,2) {$H^0_a{}$};
\node (c) at (1.5,0) {$H^0_b{}$};
\draw (a) -- (1,1);
\draw (c) -- (1,1);
\draw (1,1) -- (b);
\end{tikzpicture} &= \frac{iv}{\sqrt{2}}\begin{pmatrix}
	Z_{1}   &   2Z_{6} \\
	2Z_{6}   &   Z_{5}
\end{pmatrix}_{ab} \ ,}
\eq{
\begin{tikzpicture}[baseline=(a.base)]
\node (a) at (0,1) {$H^0_2$};
\node (b) at (1.5,2) {$H^0_a{}$};
\node (c) at (1.5,0) {$H^0_b{}$};
\draw (a) -- (1,1);
\draw (c) -- (1,1);
\draw (1,1) -- (b);
\end{tikzpicture} &= \frac{iv}{\sqrt{2}}\begin{pmatrix}
	Z_{6}   &   Z_{3}+Z_{4} \\
	Z_{3}+Z_{4}   &   Z_{7}
\end{pmatrix}_{ab} \ ,}
\eq{
\begin{tikzpicture}[baseline=(o.base),scale=1.]
\node (a) at (0,2) {$H^+_a$};
\node (b) at (0,0) {$H^0_b$};
\node (c) at (2,2) {$H^+_c$};
\node (d) at (2,0) {$H^0_d{}$};
\coordinate (o) at (1,1);
\draw[charged] (a) -- (o);
\draw[charged] (o) -- (c);
\draw (b) -- (o) -- (d);
\end{tikzpicture} &= i\begin{pmatrix}
	Z_1	&	Z_6	&	Z_6	&	Z_5	\\
	Z_6	&	Z_3	&	Z_4	&	Z_7	\\
	Z_6	&	Z_4	&	Z_3	&	Z_7	\\
	Z_5	&	Z_7	&	Z_7	&	Z_2
\end{pmatrix}_{ab,cd}\ .\label{eq:4ptsup}
}

Applying Eqs.~(\ref{eq:id_equiv1}-\ref{eq:id_equiv3}) to the four-point coupling in Eq.~(\ref{eq:4ptsup}) we arrive at
\eq{\label{eq:4ptsol}
&Z_1 + Z_2 = 2Z_3\ , \\
&Z_4 = Z_5 = 0\ ,\\
&Z_6 = Z_7 \ .
}
{Using the relations} in Eq.~(\ref{eq:4ptsol}), the $s/u$-channel amplitudes are, 
\eq{
M^s_1 = \begin{pmatrix}
    Z_{1}^2 & Z_{1}Z_{6} & Z_{1}Z_{6} & 0 \\
    Z_{1}Z_{6} & Z_{6}^2 & Z_{6}^2 & 0 \\
    Z_{1}Z_{6} & Z_{6}^2 & Z_{6}^2 & 0 \\
    0 & 0 & 0 & 0
\end{pmatrix} \ , }
\eq{
M^s_2 = \begin{pmatrix}
    Z_{6}^2 & 0 & Z_{3}Z_{6} & Z_{6}^2 \\
    0 & 0 & 0 & 0 \\
    Z_{3}Z_{6} & 0 & Z_{3}^2 & Z_{3}Z_{6} \\
    Z_{6}^2 & 0 & Z_{3}Z_{6} & Z_{6}^2
\end{pmatrix} \ , }
\eq{
M^u_1 = \begin{pmatrix}
    Z_{1}^2 & Z_{1}Z_{6} & Z_{1}Z_{6} & Z_{6}^2 \\
    Z_{1}Z_{6} & Z_{6}^2 & 0 & 0 \\
    Z_{1}Z_{6} & 0 & Z_{6}^2 & 0 \\
    Z_{6}^2 & 0 & 0 & 0
\end{pmatrix} \ , }
\eq{
M^u_2 = \begin{pmatrix}
    Z_{6}^2 & 0 & Z_{3}Z_{6} & 0 \\
    0 & 0 & Z_{6}^2 & 0 \\
    Z_{3}Z_{6} & Z_{6}^2 & Z_{3}^2 & Z_{3}Z_{6} \\
    0 & 0 & Z_{3}Z_{6} & Z_{6}^2
\end{pmatrix} \ , 
}
{
The condition $M_{11,22}=M_{12,21}=0$ then requires 
\eq{\label{eq:z6van}
Z_6=0\ .
} 
The resulting amplitude in the $t$-channel is:
\eqs{
M^t_1 &= \left(
\begin{array}{cccc}
 8 Z _1^2 s_{\tilde{\alpha}}^2 & -2 Z _1 Z _3 c_{\tilde{\alpha}} s_{\tilde{\alpha}} & 0 & 0 \\
- 2 Z _1 Z _3 c_{\tilde{\alpha}} s_{\tilde{\alpha}} & 4 Z _1 Z _3 s_{\tilde{\alpha}}^2 & 0 & 0 \\
 0 & 0 & 8 Z _1 Z _3 s_{\tilde{\alpha}}^2 &- 2 Z _3^2 c_{\tilde{\alpha}} s_{\tilde{\alpha}} \\
 0 & 0 & -2 Z _3^2 c_{\tilde{\alpha}} s_{\tilde{\alpha}} & 4 Z _3^2 s_{\tilde{\alpha}}^2 \\
\end{array}
\right) \ , \\
M^t_2 &= \left(
\begin{array}{cccc}
 8 Z _1^2 c_{\tilde{\alpha}}^2 & 2 Z _1 Z _3 c_{\tilde{\alpha}} s_{\tilde{\alpha}} & 0 & 0 \\
 2 Z _1 Z _3 c_{\tilde{\alpha}} s_{\tilde{\alpha}} & 4 Z _1 Z _3 c_{\tilde{\alpha}}^2 & 0 & 0 \\
 0 & 0 & 8 Z _1 Z _3 c_{\tilde{\alpha}}^2 & 2 Z _3^2 c_{\tilde{\alpha}} s_{\tilde{\alpha}} \\
 0 & 0 & 2 Z _3^2 c_{\tilde{\alpha}} s_{\tilde{\alpha}} & 4 Z _3^2 c_{\tilde{\alpha}}^2 \\
\end{array}
\right) \ , \\
M^t_3 &= M^t_4 = 0 \ .
} %
Solving for $M_{11,12}=M_{21,22}$ we get $Z_1=Z_3$ and, together with Eqs.~(\ref{eq:4ptsol}) and (\ref{eq:z6van}), {this leads} to the first half of Eq.~(\ref{eq:entsupp}),
\eq{
Z_1=Z_2=Z_3 = {Z}\ , \qquad Z_i=0 \ \  {\rm for}\ \  i\neq 1,2,3 \ .
} 
Then both $M^t_i$, $i=1,2$, satisfy the entanglement suppression conditions in Eqs.~(\ref{eq:id_equiv1}-\ref{eq:id_equiv3}), including the diagonal condition in Eq.~(\ref{eq:id_equiv1}): $M_{11,11}-M_{12,12}-M_{21,21}+M_{22,22}=0$. 
However, for $M^{s,u}$ we now have
\eqs{
    M^s_1=M^u_1=\begin{pmatrix} {Z}^2 & & & \\ & 0 & & \\ & & 0 & \\ & & & 0 \end{pmatrix} \ ,\quad
    M^s_2=M^u_2=\begin{pmatrix} 0 & & & \\ & 0 & & \\ & & {Z}^2 & \\ & & & 0 \end{pmatrix} \ ,
}
which do not individually satisfy  the diagonal condition. The only solution is then to require $P_{s,1} = P_{s,2}$ and $P_{u,1} = P_{u,2}$ so that the sum of the two matrices above satisfies the diagonal condition. This requires the two charged scalars to be degenerate in mass, $m_{H^\pm}=m_{G^\pm}=0$. Given the mass matrix of the charged scalars in eq.~\eqref{eq:masq}, it implies that $Y_2=-Z v^2/2=Y_1$ and all scalars other than $h$ are massless Goldstone bosons, which lead to the $SO(8)$ symmetric potential.

\end{document}